\documentclass[%
 reprint,
superscriptaddress,
 amsmath,amssymb,
 aps,
pra,
]{revtex4-2}
\usepackage{url}
\usepackage{amsfonts,amsmath, amsthm, amscd,amssymb,amsbsy,amstext,amsopn,amsxtra}
\usepackage{hyperref}
\usepackage{float}  

\usepackage{braket}
\usepackage{mathrsfs}
\usepackage{graphicx}
 \usepackage{xcolor}
\usepackage{dcolumn}
\usepackage{bm}

\begin{document}


\title{Quantum Geometry in Hyperbolic Band Theory}

\author{Ahmed Adel Mahmoud}
\email{ahmed.mahmoud@usask.ca}
\affiliation{Centre for Quantum Topology and Its Applications (quanTA), University of Saskatchewan, Saskatoon, SK, Canada}
\affiliation{Department of Mathematics and Statistics, University of Saskatchewan, Saskatoon, SK, Canada}

\author{Riccardo Sorbello}
\affiliation{Institute for Theoretical Physics and Astrophysics, Julius-Maximilians-Universit\"at W\"urzburg, W\"urzburg, Germany}

\author{Ronny Thomale}
\thanks{These authors jointly supervised this work.}
\affiliation{Institute for Theoretical Physics and Astrophysics, Julius-Maximilians-Universit\"at W\"urzburg, W\"urzburg, Germany}

\author{Steven Rayan}
\thanks{These authors jointly supervised this work.}
\affiliation{Centre for Quantum Topology and Its Applications (quanTA), University of Saskatchewan, Saskatoon, SK, Canada}
\affiliation{Department of Mathematics and Statistics, University of Saskatchewan, Saskatoon, SK, Canada}

\begin{abstract}
Hyperbolic band theory (HBT) has recently unveiled a wealth of exotic physical phenomena in negatively curved spaces. Concurrently, the quantum metric has emerged as a fundamental tensor governing quantum geometry and topological phases. In this Letter, we bridge these two frontiers by investigating the quantum metric in HBT. Because conventional Euclidean metric extraction protocols fundamentally fail in these non-commutative geometries, we introduce holonomy shaking—a novel, experimentally viable technique utilizing periodic lattice driving to dynamically extract the metric. Using the $\{8,3\}$ regular hyperbolic and kagome-like lattices as concrete examples, we demonstrate the high-fidelity dynamical extraction of the quantum metric. Our results establish a robust blueprint for probing hyperbolic quantum geometries, paving the way for the discovery of exotic topological phenomena in hyperbolic quantum matter.

\end{abstract}

\maketitle

\emph{Introduction---}The geometric structure of quantum states provides a foundational framework for modern condensed-matter physics. 
While the imaginary part of the quantum geometric tensor (QGT), the Berry curvature, has long served as a unifying language for topological phenomena such as the integer, fractional, and anomalous quantum Hall effects \cite{klitzing1980new, tsui1982two,laughlin1983anomalous, haldane1988model}, the physical significance of its real part, the quantum metric, has only recently come to the forefront \cite{torma2023essay,verma2026quantum}. 
The quantum metric is now known to critically govern macroscopic observable phenomena, including tensor monopoles \cite{tensor_monopoles1, tensor_monopoles2}, non-Hermitian physics \cite{Non-hermitian1,Non-hermitian2,Non-hermitian1n3}, and superfluidity in flat-band superconductors \cite{peotta2015superfluidity,huhtinen2022revisiting, yu2025quantum, superfluid1, superfluid2}. 
This surging interest has catalyzed the design of dynamic and spectroscopic protocols to extract the metric across various experimental platforms \cite{ozawa2018extracting,ding2022extracting,verma2025framework}. 
Quantum geometry in mechanical systems and materials continues to be a topic of interest, as evidenced by the recent work \cite{stern2026quantum}. While that work scrutinizes the quantum nature of various geometric features — in particular, they view the quantum metric as purely classical data — our present work establishes that a broad range of materials and metamaterials possess a quantum metric that contains intrinsic quantum data yet is independent of the quantum phase space (i.e. only sees the parameter space).

In a parallel frontier, hyperbolic lattices have emerged as a versatile platform for exploring the interplay of geometry, topology, and quantum dynamics. Hyperbolic band theory enables Bloch-like descriptions of particles on periodic hyperbolic lattices \cite{maciejko2021hyperbolic,maciejko2022automorphic}, motivating studies of non-Euclidean analogues of crystalline phenomena with distinctive spectral, topological, and transport properties induced by negative curvature \cite{liu2022chern,tummuru2024hyperbolic,lenggenhager2025hyperbolic,dusel2025chiral,shankar2024hyperbolic,chen2024anderson,chen2023symmetry}. 
These models have been realized experimentally in electrical circuits, photonic platforms, superconducting devices, and other synthetic quantum systems \cite{lenggenhager2022simulating,zhang2022observation,chen2023hyperbolic,zhang2023superconducting,huang2024hyperbolic,xu2025scalable}. 
Hyperbolic geometries also have important applications in quantum information processing, particularly in quantum error-correcting codes with efficient scaling properties \cite{breuckmann2016constructions, higgott2024constructions, mahmoud2025systematic,mahmoud2026hyperbolic}.

Despite these advances, the quantum geometry of hyperbolic Bloch states remains largely unexplored. Although the quantum metric has recently been studied in more general curved-space settings \cite{oancea2026quantum}, its formulation and dynamical detection within HBT have not yet been systematically developed. Two fundamental hurdles prevent progress. Firstly, formulating the metric directly requires navigating a profound dimensional mismatch: the two-dimensional real-space lattice is parameterized by a $2g$-dimensional $\mathbf{k}$-space, rendering conventional global Fourier transforms inapplicable. 
Secondly, attempting to directly adapt Euclidean metric-extraction protocols fails, as these conventional methods rely on real-space position operators whose standard interpretations collapse in non-commutative hyperbolic geometries \cite{ozawa2018extracting}. 

In this Letter, we overcome these barriers by formulating and dynamically extracting the quantum metric for the Abelian $U(1)$ sector of HBT. 
Although the physical lattice is negatively curved, its Abelian Bloch states are parameterized by the $2g$-dimensional flat Jacobian torus. Its trivial background connection eliminates the additional curvature contribution arising in general curved parameter spaces, allowing the quantum metric to be defined through ordinary derivatives with respect to the boundary-holonomy angles. This construction also resolves the dimensional mismatch between the two-dimensional real-space lattice and its $2g$-dimensional $\mathbf{k}$-space by identifying the relevant coordinates with Aharonov-Bohm holonomies around the non-contractible cycles of the quotient surface. 
Building upon this, we introduce holonomy shaking—a novel, experimentally viable spectroscopic protocol that leverages parameter-dependent holonomies on finite periodic clusters to dynamically probe the hyperbolic quantum metric. 
By validating this protocol on Abelian $\{8,3\}$ lattice clusters, we establish a concrete blueprint for exploring hyperbolic quantum geometry, laying the groundwork for investigating metric-induced phenomena, including superfluidity of interacting flat-band hyperbolic systems.

\emph{Quantum Geometry and Hyperbolic Band Theory---}In standard Euclidean band theory, translation symmetry is governed by an Abelian group, and the Brillouin zone is a $2$-torus formed by the Bloch phase factors. However, a particle hopping on a negatively curved hyperbolic lattice is governed by a non-Abelian Fuchsian translation group $\Gamma$. Upon imposing periodic boundary conditions, the compactified unit cell forms a genus-$g$ Riemann surface $\Sigma_g$ possessing $2g$ non-contractible cycles.
The application of Bloch's theorem to this non-Euclidean geometry relies on $U(1)$ representations of $\Gamma$. 
These representations are characterized by $2g$ Aharonov-Bohm (or Peierls) phases $k_j \in [0,2\pi)$, which correspond to magnetic fluxes threaded through the non-contractible cycles. Collecting these phase factors forms a $2g$-dimensional torus, the Jacobian $\text{Jac}(\Sigma_g) \cong T^{2g}$, which serves as the smooth Abelian momentum space for the hyperbolic lattice \cite{maciejko2021hyperbolic}.

In conventional Euclidean systems, the QGT is derived from the Fubini-Study distance between infinitesimally separated Bloch states \cite{provost1980riemannian}. It decomposes into the real quantum metric and the imaginary Berry curvature \cite{berry1984quantal, Berry_curvature2, ma2010abelian}. 
However, extending this geometry to intrinsically curved parameter spaces requires a generalized sub-bundle formalism \cite{oancea2026quantum}.
The QGT is constructed directly from the shape operator, substituting the standard partial derivative with a covariant derivative. 
Crucially, decomposing this generalized QGT yields a fundamental departure from the flat parameter-space limit:
\begin{equation}
Q_{AB\mu\nu} = G_{AB\mu\nu} - \frac{i}{2}\mathscr{F}_{AB\mu\nu} - \frac{1}{2}\Omega_{AB\mu\nu}.
\end{equation}
Here, $G_{AB\mu\nu}$ is the quantum metric, $\mathscr{F}_{AB\mu\nu}$ is the standard Berry curvature, and $\Omega_{AB\mu\nu}$ is an emergent curvature contribution arising directly from the non-flat background connection. 
In conventional flat condensed-matter systems, this background curvature vanishes. 
On the other hand, in hyperbolic materials, the inherent negative curvature actively modifies the quantum phase-space tensors, governing exotic transport phenomena such as spin-gravity coupling \cite{andersson2023spin}.

\begin{figure}
    \centering
    \includegraphics{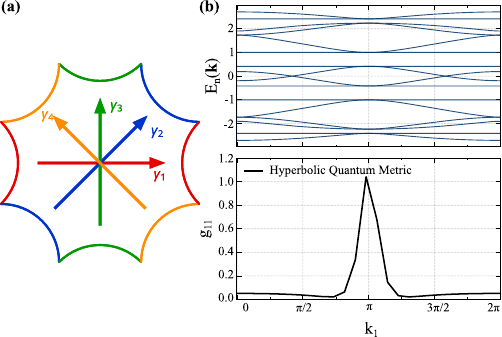}
    \caption{(a) The octagonal fundamental domain used for the $\{8,3\}$ hyperbolic lattice. The colored arrows denote the non-contractible generators of the Fuchsian translation group $\Gamma$, mapping opposite edges to establish periodic boundary conditions. (b) The Bloch spectrum and the isolated lowest-band quantum metric along the trajectory $\mathbf{k}=(k_1,\pi,\pi,\pi)$. At the high-symmetry torsion point $k_1=\pi$, the bands touch, causing the rank-one quantum metric to develop a sharp singularity.}
    \label{fig:hyperbolic_geometry_and_metric}
\end{figure}

To define a physically extractable quantum metric, one must carefully delineate which portion of the Hilbert space is being probed. The full quantum state counting on a hyperbolic lattice encompasses an immense, largely intractable spectrum of higher-rank, non-Abelian group representations. Hyperbolic band theory provides a robust mathematical foothold by applying Bloch's theorem strictly to the one-dimensional $U(1)$ representations of $\Gamma$. It is crucial to emphasize that this framework is not a coarse approximation of the overall energy spectrum; rather, it constitutes an exact, algebraically closed subset of states known as the Abelian sector.

Crucially, restricting our analysis to finite Abelian clusters reveals an elegant geometric simplification: while the physical real-space lattice is intrinsically negatively curved, the corresponding higher-dimensional parameter space (the Jacobian torus) is globally flat. Hence, the emergent curvature contribution $\Omega_{AB\mu\nu}$ naturally vanishes and the covariant derivatives within the generalized sub-bundle formalism reduce exactly to conventional partial derivatives. The QGT, and by extension the quantum metric, can therefore be rigorously defined and evaluated using standard $\mathbf{k}$-space derivatives $\partial_{k_\mu}$. For a Bloch state $\vert{}u(\mathbf{k})\rangle$ belonging to an isolated band, the quantum metric reduces to its conventional formulation:
\begin{equation}
    g_{\mu\nu}(\mathbf{k}) = \text{Re} \langle \partial_\mu u(\mathbf{k}) \vert{} \left(1 - \vert{}u(\mathbf{k})\rangle\langle u(\mathbf{k})\vert{}\right) \vert{} \partial_\nu u(\mathbf{k}) \rangle.
    \label{eqn: quantum metric}
\end{equation}
This formulation completely bypasses the geometric complexities of the underlying spatial curvature. We explicitly verify this profound dimensional and geometric decoupling in the subsequent sections through a high-fidelity kinematic metric extraction in $\mathbf{k}$-space.

\emph{High-Symmetry Points and Metric Singularities---}The Abelian hyperbolic quantum metric of the $\{8,3\}$ lattice exhibits a nontrivial structure over the Jacobian \(T^{4}\). Using the full 96-element symmetry group of the Bolza surface, we identify torsion points at which symmetry enforces Bloch-band degeneracies (see Appendix D for the complete orbit classification). 
As such a degeneracy is approached, the squared interband energy gap in the spectral representation vanishes; when the corresponding matrix elements remain nonzero, the isolated-band metric diverges. 
At the degeneracy itself, the rank-one projector, and hence the isolated-band metric, becomes ill-defined. Figure~\ref{fig:hyperbolic_geometry_and_metric}(b) shows this singular enhancement along \(\mathbf{k}=(k_1,\pi,\pi,\pi)\) near the 2-torsion point \((\pi,\pi,\pi,\pi)\), demonstrating how the symmetries of the Bolza surface constrain the quantum geometry.

\emph{Kinematic Metric Extraction in Momentum Space---}To systematically probe hyperbolic quantum geometry and bridge it with dynamic observables, we exploit the flat $\mathbf{k}$-space structure of Abelian clusters. In this case, the current operator is globally well-defined by the standard $\mathbf{k}$-derivative, $V_j(\mathbf{k}) = \partial_{k_j} H(\mathbf{k})$. This robust formulation allows us to engineer a direct kinematic analogue to Euclidean lattice shaking in order to dynamically probe the metric. We modulate the $j$-th Aharonov-Bohm phase via a time-periodic harmonic drive with applied force $F(t) = 2E\cos(\omega t)$. The corresponding vector potential is $A(t) = (2E /\omega)\sin(\omega t)$, and the dynamically modulated Hamiltonian is therefore:

\begin{equation}
\begin{aligned}
H(\mathbf{k}, t) &= H\left(\mathbf{k} + \frac{2E}{\omega}\sin(\omega t)\mathbf{e}_j\right), \\ 
&\approx H(\mathbf{k}) + \frac{2E}{\omega}\sin(\omega t)\partial_{k_j}H(\mathbf{k}),
\end{aligned}
\end{equation}
where we use units in which \(\hbar=1\) throughout.

\begin{figure*}[t]
\centering
\includegraphics{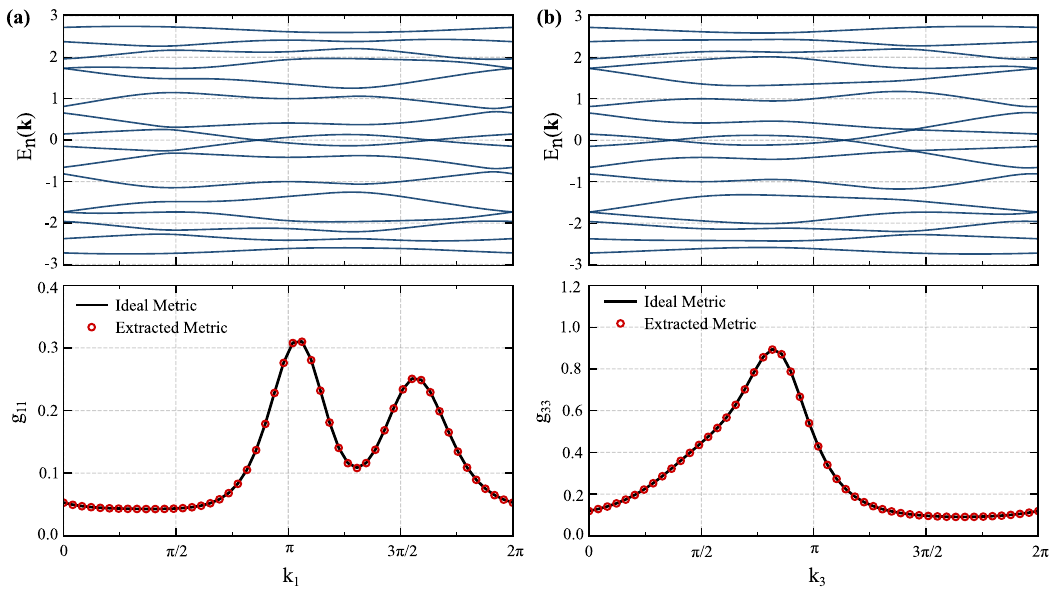}
\caption{Kinematic extraction of the diagonal quantum metric components in the hyperbolic Abelian momentum space. (a) The energy spectrum and the highly accurate extraction of $g_{11}$ (red circles) against the exact analytical formula (solid black line) along $k_1 \in [0, 2\pi]$ with $k_2=\pi, k_3=k_4=\pi/2$. (b) The corresponding spectrum and precision extraction for the $g_{33}$ component along $k_3 \in [0, 2\pi]$ with $k_1=k_2=\pi, k_4=\pi/2$.}
\label{fig:kinematic_extraction}
\end{figure*}

By tracking the excitation rates between bands induced by this modulation, we can extract the quantum metric dynamically. To verify the soundness of this kinematic $\mathbf{k}$-space formulation before adapting it to the real-space constraints of finite Abelian clusters, we execute the extraction protocol numerically. As shown in Fig.~\ref{fig:kinematic_extraction}, the metric values extracted dynamically via the Aharonov-Bohm phase modulation agree closely with the metric evaluated directly from the eigenstates.
This agreement verifies that modulating the $j$-th Abelian holonomy couples through $\partial_{k_j}H(\mathbf{k})$ and thereby probes the corresponding quantum-metric component $g_{jj}$ in the Abelian sector of HBT.

\emph{Extracting the Hyperbolic Quantum Metric Through Holonomy Shaking ---}We now formulate a real-space, dynamic, periodic-driving protocol to probe the quantum metric of hyperbolic tight-binding models \cite{boettcher2022crystallography}. 
In Euclidean lattices, the metric is routinely extracted by applying a periodic force proportional to the real-space position operator \cite{ozawa2018extracting}. 
However, a direct generalization to hyperbolic space encounters a fundamental geometric obstruction: hyperbolic crystal momentum is not dual to a local coordinate on the Poincaré disk, but rather to the homology directions of the compact quotient surface. 
Consequently, formal real-space position operators fail to descend to globally defined, diagonal operators on finite periodic clusters (see Appendix A for an algebraic proof). 
To circumvent this barrier, we propose holonomy shaking, an experimentally viable protocol expressed through real-space hopping phases and their associated bond-current operators.

We first construct a holonomy-dependent Hamiltonian. 
Let $\Gamma$ denote the Fuchsian translation group of the infinite hyperbolic lattice, $\Gamma_{PBC} \triangleleft \Gamma$ denote a normal subgroup, and $Q = \Gamma/\Gamma_{PBC}$ denote the finite Abelian quotient group defining translational symmetries of the periodic Abelian cluster.
Each hopping process $\eta \in \Gamma$ possesses an Abelianization vector $b(\eta) \in \mathbb{Z}^{2g}$ specifying its homology class. 
We introduce a continuous family of Abelian holonomies $\boldsymbol{\theta} = (\theta_1, \ldots, \theta_{2g})$ by attaching Peierls phases to the homological displacement of each hopping:
\begin{equation}
    \hat{H}_{\text{real}}(\boldsymbol{\theta}) = \sum_{\eta} e^{i \boldsymbol{\theta} \cdot b(\eta)} \hat{R}_{\eta} \otimes h_{\eta},
    \label{eqn: holonomy-defined hamiltonian}
\end{equation}
where $\hat{R}_{\eta}$ generates left translations on the quotient, and $h_{\eta}$ represents intra-cell connections. 
The operator for infinitesimal holonomy variations is therefore defined as:
\begin{equation}
    \hat{J}_j^{\text{real}} = \left. \frac{\partial \hat{H}_{\text{real}}(\boldsymbol{\theta})}{\partial \theta_j} \right\vert_{\boldsymbol{\theta}=\mathbf{0}} = i \sum_{\eta} b_j(\eta) \hat{R}_{\eta} \otimes h_{\eta}.
    \label{eqn: J^real formula}
\end{equation}
Crucially, $\hat{J}_j^{\text{real}}$ acts as a globally well-defined, weighted sum of bond-current operators. 
Its weights are determined exclusively by the Abelianized homology classes of the hopping processes, entirely bypassing the need for a local position operator.

We dynamically probe the quantum metric by introducing a time-dependent modulation to these boundary holonomies. 
Analogous to Euclidean linear lattice shaking, driving the $j$-th homology direction with a phase modulation $\theta_j(t) = (2E/\omega)\sin(\omega t)$ yields the time-dependent real-space Hamiltonian:
\begin{equation}
    \hat{H}^{(j)}(t) = \hat{H}_{\textrm{real}} + \frac{2E}{\omega}\sin(\omega t)\hat{J}_j^{\textrm{real}} + \mathcal{O}\left( \frac{E^2}{\omega^2} \right).
\end{equation}
Experimentally, this driving corresponds to a periodic modulation of the complex hopping amplitudes, $t_e \to t_e e^{i\theta_j(t)m_{e,j}}$. 
The required laboratory control is therefore a local modulation of bond phases rather than a direct manipulation of hyperbolic crystal momentum.
Such dynamic phase modulation, effectively mimicking a synthetic, time-varying Peierls gauge field, is readily achievable in modern synthetic matter platforms. For instance, the complex phases can be engineered dynamically via time-modulated tunable admittances in topolectrical circuits \cite{chen2023hyperbolic}, or the parametric driving of tunable couplers in superconducting qubit arrays \cite{roushan2017chiral}, or via electro-optic modulation in photonic resonator lattices \cite{dutt2019experimental, yuan2021tutorial}. Holonomy shaking thus provides a direct experimental probe of the quantum metric for the Abelian sector of HBT.

Starting from the ground state of the Bloch Hamiltonian at $\mathbf{k}_{\kappa}$, we measure the population $P_{\mathrm{exc},j}(\mathbf{k}_{\kappa};\omega,T)$ transferred to excited bands after time $T$, using the undriven band projector. Defining $\Gamma_{j,T}=P_{\mathrm{exc},j}/T$, the lowest-band metric is obtained as
\begin{equation}
    g_{jj}^{(0)}(\mathbf{k}_{\kappa})
    \simeq \frac{1}{4\pi E^2}
    \int_0^{\infty} d\omega\,
    \Gamma_{j,T}(\mathbf{k}_{\kappa};\omega).
    \label{eq:finite_time_metric_extraction}
\end{equation}
The relation assumes weak driving and observation times long compared with the inverse interband gaps. The $4\pi$ normalization retains the low-frequency contribution of the finite-time sine drive; Appendix E derives this result and explains its distinction from the resonant-rate formula~\cite{ozawa2018extracting}.

\begin{figure*}[t]
\centering
\includegraphics[]{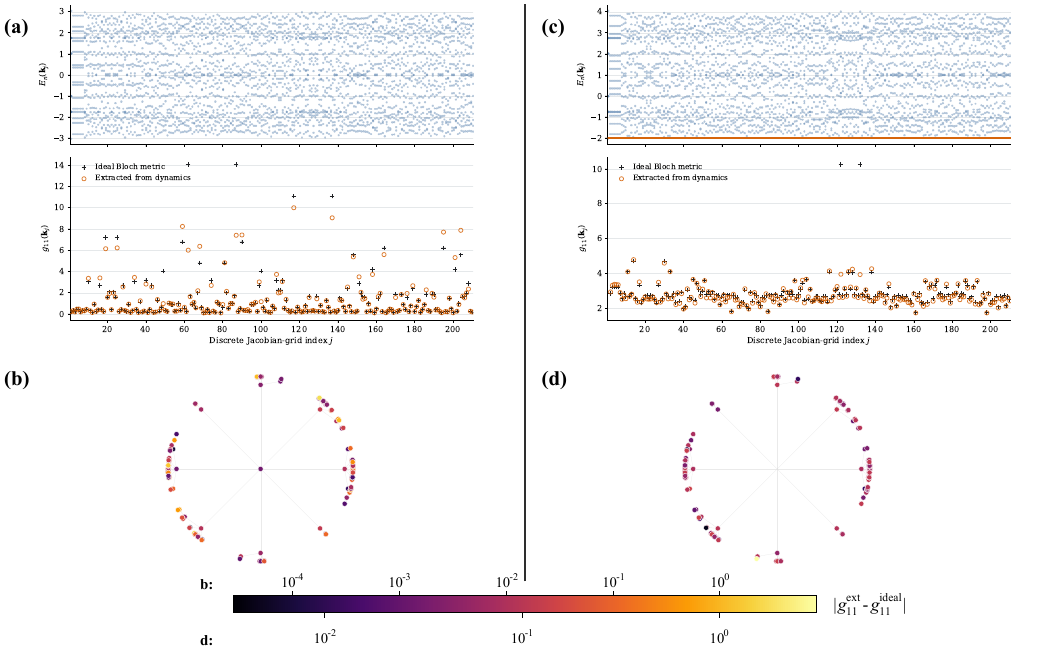}
\caption{Quantum-metric extraction on a 210-cell Abelian cluster.
\textbf{(a)} Spectrum of the regular hyperbolic $\{8,3\}$ lattice and its Abelian metric component $g_{11}$, comparing the ideal values (black crosses) with those extracted through holonomy shaking (orange circles).
\textbf{(b)} Logarithmic spatial map of the absolute extraction error $|g_{11}^{\mathrm{ext}}-g_{11}^{\mathrm{ideal}}|$ on the underlying lattice with one metric value assigned to each unit cell.
\textbf{(c)} Spectrum of the $\{8,3\}$ Kagome-like lattice, prominently exhibiting an eightfold-degenerate flat ground-state band, together with the extracted trace $\operatorname{Tr}g_{11}$ of its non-Abelian quantum metric.
\textbf{(d)} Spatial map of the absolute error in $\operatorname{Tr}g_{11}$.
In \textbf{(b)} and \textbf{(d)}, the symmetry-related high-error regions reflect the underlying hyperbolic translation symmetry and identify points where the relevant ground-state band or degenerate manifold approaches higher bands. 
The resulting metric enhancement and reduced spectral isolation make the finite-time linear-response extraction least reliable, demonstrating that the error is governed by the physical band geometry rather than uncorrelated numerical noise.
}
\label{fig:metric_real_space}
\end{figure*}

\emph{Results---}We implement the holonomy-shaking protocol on two hyperbolic tight-binding models defined on the same finite Abelian cluster: the regular \(\{8,3\}\) lattice and its Kagome-like counterpart. Periodic boundary conditions are imposed through a finite-index normal subgroup of the Fuchsian translation group, yielding an Abelian quotient of 210 unit cells. Its 210 one-dimensional \(U(1)\) representations discretize the four-dimensional Jacobian torus into 210 allowed momenta \(\mathbf{k}_j\) (see Appendix B for details of the quotient group structure).

The Abelian translation group should be distinguished from the gauge structure of the energy eigenspaces. It allows the boundary holonomies to be parameterized by \(U(1)\) phases and interpreted as generalized momenta, but does not require the energy bands to be nondegenerate. The regular \(\{8,3\}\) model has a generically nondegenerate lowest band described by a rank-one Abelian quantum metric. By contrast, the Kagome-like model has an eightfold-degenerate flat-band ground state whose projector defines a rank-eight vector bundle with a non-Abelian quantum metric.

At each allowed momentum, we periodically modulate the real-space hopping phases along the Abelian holonomy direction \(\theta_1\). The induced transitions from the ground state to the excited spectrum are frequency-integrated within the linear-response regime. For the regular lattice, this response yields the metric component \(g_{11}\); for the Kagome-like lattice, summing over the ground-state manifold yields the gauge-invariant trace \(\operatorname{Tr}g_{11}\) of the non-Abelian metric.

Figure~\ref{fig:metric_real_space}(a) compares the ideal \(g_{11}\) of the regular \(\{8,3\}\) lattice with its dynamically extracted values, while Fig.~\ref{fig:metric_real_space}(b) shows the spatial distribution of the residual error. The results are plotted against a scalar index \(j\) enumerating the 210 discrete momenta. The extraction reproduces the ideal metric with high fidelity across most of the discretized Jacobian, with visible discrepancies near band degeneracies. As shown in Fig.~\ref{fig:metric_real_space}(b), the high-error regions form symmetry-related orbits and cluster near torsion momenta where the lowest band approaches or touches the excited spectrum. There, the protecting gap narrows or closes, enhancing or rendering singular the isolated-band metric and reducing the accuracy of the finite-time perturbative extraction.

Figures~\ref{fig:metric_real_space}(c) and (d) present the corresponding results for the Kagome-like lattice. Unlike the regular \(\{8,3\}\) model, this system exhibits an eightfold-degenerate flat-band ground state and therefore requires a projector-valued, non-Abelian metric. 
The exact touching point at the origin is excluded because the fixed-rank ground-state projector is ill-defined there.
The error map in Fig.~\ref{fig:metric_real_space}(d) again exhibits a pronounced symmetry-related structure, with the largest deviations occurring where the eight-dimensional ground-state manifold is least isolated from the excited bands. Internal degeneracy within the flat band does not invalidate the protocol because the manifold is treated collectively through its projector. Instead, the accuracy is controlled by the gap separating the entire manifold from the remaining spectrum. The maps in Figs.~\ref{fig:metric_real_space}(b) and \ref{fig:metric_real_space}(d) thus reveal the group-theoretic organization of the exceptional momenta and show that the extraction error is governed by the physical band geometry rather than numerical noise.

\emph{Conclusion---}Our results establish holonomy shaking as a unified probe of both rank-one Abelian and higher-rank non-Abelian quantum geometries of the Abelian sector of HBT. In particular, the Kagome-like model demonstrates that an Abelian hyperbolic \(\mathbf{k}\)-space construction can support and resolve a non-Abelian quantum metric whenever the relevant band subspace is degenerate. 

Two directions follow naturally. First, introducing Hubbard-type interactions into the hyperbolic Kagome-like flat band would enable investigation of the geometric contribution to the superfluid weight and of interaction-driven phases stabilized by the combination of flat dispersion, non-Abelian quantum geometry, and negative curvature.
Second, extending holonomy shaking to non-Abelian clusters would probe higher-dimensional irreducible representations of the Fuchsian translation group that are absent from the present $U(1)$ description. This extension necessitates deviations from the present flat Abelian Jacobian considered here.
Hence, their quantum metric could reveal geometric features inaccessible in the Abelian sector, potentially unlocking rich topological and observable phenomena, thereby advancing a complete geometric band theory of hyperbolic matter.

\emph{Acknowledgments---}This project was initiated during a research stay at Julius-Maximilians-Universität Würzburg by A.A.M., who gratefully acknowledges financial support from the Mitacs Globalink Research Award. S.R. and A.A.M. were supported by the NSERC Discovery Grant and PIMS Site Director Support Grant (both of S.R.). R.S. and R.T. acknowledge the support from the W\"urzburg-Dresden Cluster of Excellence on Complexity, Topology and Dynamics in Quantum Matter – ctd.qmat, Project-ID 390858490 – EXC 2147.

\emph{Data availability---}The numerical data underlying Figs.~\ref{fig:kinematic_extraction} and~\ref{fig:metric_real_space}, including the extraction parameters and the coset table defining the 210-cell Abelian cluster, are openly available in the Zenodo repository~\cite{Mahmoud2026QuantumGeometryData}.

\bibliography{apssamp}

\section*{End Matter}

\emph{Appendix A: Algebraic Formalism of Abelianization and the Descent Obstruction ---}Let $\Gamma$ be the Fuchsian translation group of the hyperbolic lattice, and let $\Sigma_g = \mathbb{D}/\Gamma$ denote the compact quotient surface of genus $g$. 
The Abelianization of $\Gamma$ is given by the quotient group
\begin{equation}
    \Gamma_{\mathrm{ab}} = \Gamma/[\Gamma,\Gamma] \cong H_1(\Sigma_g,\mathbb{Z}) \cong \mathbb{Z}^{2g}.
\end{equation}
By choosing an integral basis for the first homology group $H_1(\Sigma_g,\mathbb{Z})$, we can define the Abelianization map as
\begin{equation}
    b: \Gamma \longrightarrow \mathbb{Z}^{2g}, \qquad b(\gamma) = \left( b_1(\gamma),\ldots,b_{2g}(\gamma) \right).
\end{equation}
Since $b$ is a group homomorphism, it satisfies $b(\eta \gamma) = b(\eta) + b(\gamma)$. As an illustrative example, consider a lattice with a genus $g=2$ translation group, such as the regular $\{8,8\}$ tessellation. 
The defining relation of its Fuchsian group possesses a vanishing exponent sum for each of the four generators. One may therefore naturally assign $b(\gamma_1) = (1,0,0,0)$, $b(\gamma_2) = (0,1,0,0)$, $b(\gamma_3) = (0,0,1,0)$, and $b(\gamma_4) = (0,0,0,1)$. 
For any arbitrary word in these generators, $b(\gamma)$ is obtained by summing the signed exponent of each respective generator. Consequently, all commutators are mapped identically to zero; for instance, $b(\gamma_1\gamma_2\gamma_1^{-1}\gamma_2^{-1}) = (0,0,0,0)$. 
Importantly, the Abelianization map is not generally injective. Distinct elements of $\Gamma$ may possess identical Abelianization vectors. For example, $b(\gamma_1\gamma_2) = b(\gamma_2\gamma_1) = (1,1,0,0)$ even though $\gamma_1\gamma_2 \neq \gamma_2\gamma_1$ in the non-Abelian group $\Gamma$. Thus, $b(\gamma)$ solely records the homology class of a translation, not the specific translated unit cell. A $U(1)$ representation of $\Gamma$ strictly factors through its Abelianization and takes the form
\begin{equation}
    \chi_{\mathbf{k}}(\gamma) = \exp\left[ i\mathbf{k} \cdot b(\gamma) \right], \qquad \mathbf{k} \in \mathbb{R}^{2g}/2\pi\mathbb{Z}^{2g}.
\end{equation}
The components $k_j$ serve as coordinates on the Abelian character torus $\mathrm{Hom}(\Gamma,U(1)) \cong \mathrm{Jac}(\Sigma_g)$. Crucially, they must be interpreted as Aharonov--Bohm phases associated with an integral homology basis of $\Sigma_g$, and not as physical momenta dual to local Euclidean coordinates on the Poincar\'e disk. On the infinite universal cover (the Poincar\'e disk), one can formally define a diagonal real-space ``winding'' operator $\hat{B}_j$ that acts on a state localized at cell $\gamma$ and orbital $\alpha$ as $\hat{B}_j \ket{\gamma,\alpha} = b_j(\gamma) \ket{\gamma,\alpha}$. However, this operator fundamentally fails to descend to finite periodic geometries. Consider a finite periodic cluster defined by a finite-index normal subgroup $\Gamma_{\mathrm{PBC}} \triangleleft \Gamma$, with the finite Abelian quotient defined as $Q = \Gamma/\Gamma_{\mathrm{PBC}}$. The unit cells of the finite lattice are labeled by the cosets $[\gamma] \in Q$. A direct restriction of the infinite-lattice winding operator to this finite cluster would require
\begin{equation}
    \hat{B}_j \ket{[\gamma],\alpha} = b_j(\gamma) \ket{[\gamma],\alpha}.
\end{equation}
However, a given coset can be equally represented by both $\gamma$ and $\gamma \beta$, where $\beta \in \Gamma_{\mathrm{PBC}}$. Because the Abelianization map is a homomorphism, $b_j(\gamma \beta) = b_j(\gamma) + b_j(\beta)$. For the operator $\hat{B}_j$ to be well-defined on the quotient space, its eigenvalue must be independent of the chosen coset representative. This enforces the strict condition
\begin{equation}
    b_j(\beta) = 0 \qquad \text{for every } \beta \in \Gamma_{\mathrm{PBC}}.
\end{equation}
This condition cannot hold for a finite-index subgroup $\Gamma_{\mathrm{PBC}}$. Indeed, choose a generator $\gamma_j$ with $b_j(\gamma_j)=1$. Since the quotient $Q$ is finite, the coset of $\gamma_j$ has finite order $r$, so $\gamma_j^r\in\Gamma_{\mathrm{PBC}}$. However, $b_j(\gamma_j^r)=r\neq0$, contradicting the condition required for descent. Thus, the winding operator $\hat{B}_j$ does not define a representative-independent diagonal operator on the finite periodic cluster. Its failure to descend therefore obstructs a globally defined diagonal position operator with this property on the finite periodic cluster. Holonomy shaking implements the required drive directly through the hopping phases.

\emph{Appendix B: Finite Quotients and Abelian Momentum Sampling---}A finite periodic cluster is specified by a finite-index normal subgroup $\Gamma_{\mathrm{PBC}}\triangleleft\Gamma$, with translation group $Q=\Gamma/\Gamma_{\mathrm{PBC}}$. The allowed representation sectors are the irreducible representations of $Q$. If $Q$ is non-Abelian, these include representations of dimension greater than one. Here, we restrict attention to Abelian quotients, whose irreducible representations are all one-dimensional characters. By the fundamental theorem of finitely generated Abelian groups, a finite Abelian group admits a decomposition into cyclic groups of prime-power order~\cite{fraleigh2003first}. Its $|Q|$ distinct characters specify the allowed Abelian momenta of the periodic cluster.

The positions of these momenta within the Jacobian depend on the quotient map from $\Gamma$ to $Q$. To express this dependence, let $b:\Gamma\to\mathbb{Z}^{2g}$ be the Abelianization map and define the finite-index sublattice $L=b(\Gamma_{\mathrm{PBC}})\subset\mathbb{Z}^{2g}$. Since $Q$ is Abelian, $\Gamma_{\mathrm{PBC}}$ contains the commutator subgroup, and $Q\cong\mathbb{Z}^{2g}/L$. A character $\chi_{\mathbf{k}}(\gamma)=e^{i\mathbf{k}\cdot b(\gamma)}$ descends to $Q$ precisely when it is trivial on $\Gamma_{\mathrm{PBC}}$. Consequently, the allowed momenta form the finite set
\begin{equation*}
\mathcal{K}_Q=
\left\{
\mathbf{k}\in\mathbb{R}^{2g}/2\pi\mathbb{Z}^{2g}:
\mathbf{k}\cdot\boldsymbol{\ell}\in2\pi\mathbb{Z}
\ \text{for every }\boldsymbol{\ell}\in L
\right\}.
\end{equation*}
Thus, the abstract group structure determines the number of allowed momenta, while the sublattice $L$ determines their arrangement within the Jacobian. The derivatives entering the quantum metric are defined through the continuous holonomy-dependent Hamiltonian and evaluated at these discrete momenta.

For the $\{8,3\}$ lattice model considered here, we construct the periodic subgroup by intersecting normal subgroups of indices $7$, $5$, $3$, and $2$~\cite{chen2024anderson}. The resulting Abelian quotient has order $210$ and admits the decomposition $Q\cong\mathbb{Z}_7\times\mathbb{Z}_5\times\mathbb{Z}_3\times\mathbb{Z}_2$. Its characters provide $210$ sampling points in the four-dimensional Jacobian. The normal subgroups were enumerated and filtered using the LINS package within GAP~\cite{LINS0.9,GAP4}.

\emph{Appendix C: Hyperbolic Wavepackets and Band-Theoretic Evolution---}A natural question is whether a localized wavepacket can be constructed on a hyperbolic lattice and subsequently analyzed using hyperbolic band theory. This is possible, provided one distinguishes between localization on the Poincar\'e disk and the Abelian character variables that label automorphic Bloch eigenstates.

Let $z_{\gamma,\alpha}\in\mathbb{D}$ denote the embedded position of orbital $\alpha$ in the unit cell labelled by $\gamma\in\Gamma$, and let $z_0$ be the wavepacket center. A geometrically localized state should be defined using the hyperbolic distance $d_{\mathbb{H}}$ on the disk, rather than the Euclidean distance of the embedding. A representative wavepacket with character $\mathbf{k}_0$ is
\begin{equation*}
\ket{\Psi_{\mathrm{raw}}}
=
\mathcal{N}
\sum_{\gamma,\alpha}
\exp\left[
i\mathbf{k}_0\cdot b(\gamma)
-
\frac{d_{\mathbb{H}}(z_{\gamma,\alpha},z_0)^2}
{2\sigma_{\mathrm{real}}^2}
\right]
v_\alpha\ket{\gamma,\alpha},
\end{equation*}
where $b:\Gamma\to\mathbb{Z}^{2g}$ is the Abelianization map, $\mathbf{k}_0\in\mathrm{Jac}(\Sigma_g)$ is an Abelian Bloch phase, $v_\alpha$ is an internal orbital spinor, and $\sigma_{\mathrm{real}}$ is a hyperbolic length scale. Thus the disk coordinates determine the real-space envelope, while the character variables determine the automorphic phase structure.

On a finite Abelian quotient, the allowed characters form a discrete set $\mathcal{K}_Q\subset\mathrm{Jac}(\Sigma_g)$. Expanding the state in hyperbolic Bloch eigenstates gives
\begin{equation}
\ket{\Psi_{\mathrm{raw}}}
=
\sum_{\mathbf{k}\in\mathcal{K}_Q}
\sum_n
c_n(\mathbf{k})\ket{\Phi_{n,\mathbf{k}}}.
\end{equation}
If the dynamics of an isolated band $n$ is desired, the state is first projected onto that band and renormalized. Its band-theoretic evolution is then
\begin{equation}
\ket{\Psi_n(t)}
=
\sum_{\mathbf{k}\in\mathcal{K}_Q}
c_n(\mathbf{k})e^{-iE_n(\mathbf{k})t}
\ket{\Phi_{n,\mathbf{k}}},
\end{equation}
with $E_n(\mathbf{k})$ obtained from the hyperbolic Bloch Hamiltonian.

In practice, projection onto an isolated band can broaden the packet, especially when $\sigma_{\mathrm{real}}$ is too small. Accordingly, $\sigma_{\mathrm{real}}$ cannot be chosen arbitrarily: it must be selected so that the normalized projected state remains localized near the intended center and its inverse participation ratio is not substantially changed by projection.

\emph{Appendix D: High-Symmetry Torsion Points---}Applying the 96-element symmetry group $G = \langle R,S,T,U \rangle$ of the Bolza surface to the Abelian Brillouin zone $\mathrm{Jac}(\Sigma_2) \simeq (\mathbb{R}/2\pi\mathbb{Z})^4$ reveals distinct orbits of high-symmetry momenta where band touchings occur. We identify a 16-point orbit corresponding to third-order torsion points ($3\mathbf{k} = \mathbf{0} \pmod{2\pi}$), given by:
\begin{equation*}
\begin{aligned}
   \mathcal{K}_3 = \{ 
   &(0, \tfrac{2\pi}{3}, \tfrac{2\pi}{3}, \tfrac{4\pi}{3}), && (0, \tfrac{2\pi}{3}, \tfrac{4\pi}{3}, \tfrac{4\pi}{3}), && (0, \tfrac{4\pi}{3}, \tfrac{2\pi}{3}, \tfrac{2\pi}{3}), \\
   &(0, \tfrac{4\pi}{3}, \tfrac{4\pi}{3}, \tfrac{2\pi}{3}), && (\tfrac{2\pi}{3}, 0, \tfrac{2\pi}{3}, \tfrac{2\pi}{3}), && (\tfrac{2\pi}{3}, 0, \tfrac{2\pi}{3}, \tfrac{4\pi}{3}), \\
   &(\tfrac{2\pi}{3}, \tfrac{2\pi}{3}, 0, \tfrac{2\pi}{3}), && (\tfrac{2\pi}{3}, \tfrac{2\pi}{3}, \tfrac{4\pi}{3}, 0), && (\tfrac{2\pi}{3}, \tfrac{4\pi}{3}, 0, \tfrac{4\pi}{3}), \\
   &(\tfrac{2\pi}{3}, \tfrac{4\pi}{3}, \tfrac{4\pi}{3}, 0), && (\tfrac{4\pi}{3}, 0, \tfrac{4\pi}{3}, \tfrac{2\pi}{3}), && (\tfrac{4\pi}{3}, 0, \tfrac{4\pi}{3}, \tfrac{4\pi}{3}), \\
   &(\tfrac{4\pi}{3}, \tfrac{2\pi}{3}, 0, \tfrac{2\pi}{3}), && (\tfrac{4\pi}{3}, \tfrac{2\pi}{3}, \tfrac{2\pi}{3}, 0), && (\tfrac{4\pi}{3}, \tfrac{4\pi}{3}, 0, \tfrac{4\pi}{3}), \\
   &(\tfrac{4\pi}{3}, \tfrac{4\pi}{3}, \tfrac{2\pi}{3}, 0) \}.
\end{aligned}
\end{equation*}
Furthermore, we identify a 3-point orbit of second-order torsion points ($2\mathbf{k} = \mathbf{0} \pmod{2\pi}$):
\begin{equation*}
    \mathcal{K}_2 = \left\{ (0,\pi,0,\pi), (\pi,0,\pi,0), (\pi,\pi,\pi,\pi) \right\}.
\end{equation*}
These correspond to time-reversal-invariant momenta where the isolated-band metric exhibits the singular divergences described in the main text, see Fig.~\ref{fig:hyperbolic_geometry_and_metric}(b).

\emph{Appendix E: Finite-Time Extraction Formula---}At fixed momentum, let $\ket{0}$ be the nondegenerate ground state of the undriven Bloch Hamiltonian $H_0$. Under $H(t)=H_0+(2E/\omega)\sin(\omega t)J_j$, with $J_j=\partial_{k_j}H_0$, the probability of occupying an undriven excited state $\ket{m}$ at time $T$ is, to leading order in $E^2$,
\begin{equation}
P_m(\omega,T)=4E^2|J_{m0}|^2
\left|\int_0^T dt\,
e^{i\Delta_m t}\frac{\sin(\omega t)}{\omega}\right|^2,
\end{equation}
where $\Delta_m=E_m-E_0>0$ and $J_{m0}=\bra{m}J_j\ket{0}$. Using
\begin{equation}
\int_0^\infty d\omega\,
\frac{\sin(\omega t)\sin(\omega s)}{\omega^2}
=\frac{\pi}{4}\bigl(t+s-|t-s|\bigr)
=\frac{\pi}{2}\min(t,s),
\end{equation}
where $t,s\geq0$ and $\min(t,s)$ denotes the smaller of the two integration times. Hence, at the same perturbative order,
\begin{equation}
\begin{aligned}
\int_0^\infty d\omega\,\Gamma_{j,T}(\omega)
&=4\pi E^2\sum_{m>0}\frac{|J_{m0}|^2}{\Delta_m^2}\\
&\quad\times\left[1-\frac{\sin(\Delta_mT)}{\Delta_mT}\right],
\end{aligned}
\end{equation}
with $\Gamma_{j,T}=\sum_{m>0}P_m/T$. Since $g_{jj}^{(0)}=\sum_{m>0}|J_{m0}|^2/\Delta_m^2$, Eq.~\eqref{eq:finite_time_metric_extraction} follows for $\Delta_mT\gg1$.

\end{document}